\documentclass[10pt]{article}
\usepackage{graphicx}

\usepackage{color}
\usepackage{cite}

\usepackage{setspace} 
\usepackage[labelfont=bf,labelsep=period,justification=raggedright]{caption}

\makeatletter
\renewcommand{\@biblabel}[1]{\quad#1.}
\makeatother

\date{}

\begin{document}

\begin{flushleft}
{\Large \textbf{Two types of well followed users in the followership networks of Twitter}
}
\\
Kodai Saito$^{1}$ and Naoki Masuda$^{1,\ast}$
\\
\bf{1} Department of Mathematical Informatics,
The University of Tokyo,
Bunkyo, Tokyo, Japan
\\
$\ast$ E-mail: masuda@mist.i.u-tokyo.ac.jp
\end{flushleft}

\section*{Abstract}
In the Twitter blogosphere, the number of followers is probably the most basic and succinct quantity for measuring popularity of users. However, the number of followers can be manipulated in various ways; we can even buy follows. Therefore, alternative popularity measures for Twitter users on the basis of, for example, users' tweets and retweets, have been developed. In the present work, we take a purely network approach to this fundamental question. First, we find that two relatively distinct types of users possessing a large number of followers exist, in particular for Japanese, Russian, and Korean users among the seven language groups that we examined. A first type of user follows a small number of other users. A second type of user follows approximately the same number of other users as the number of follows that the user receives. Then, we compare local (i.e., egocentric) followership networks around the two types of users with many followers. We show that the second type, which is presumably uninfluential users despite its large number of followers, is characterized by high link reciprocity, a large number of friends (i.e., those whom a user follows) for the followers, followers' high link reciprocity, large clustering coefficient, large fraction of the second type of users among the followers, and a small PageRank. Our network-based results support that the number of followers used alone is a misleading measure of user's popularity. We propose that the number of friends, which is simple to measure, also helps us to assess the popularity of Twitter users.

\section*{Introduction}

Twitter started to operate on July 2006 and
possessed over $5.5\times10^8$ registered users as of May 2013. Registered
users can send
and read text message of up to 140 characters called ``tweet''.
In social microblogging services including Twitter, users can follow or unfollow activities such as posting of other users of interest.
The presumably simplest indicator of the popularity of users in Twitter is
the number of followers \cite{Ghosh2012WWW}.
This quantity is shown on the profile webpage of each user, which makes it even popular. In addition, main activity-related measures of users such as the retweet rate
are known to be also proportional to the number of followers of a user \cite{Suh2010SocialCom}.

However, the number of followers may be misguiding as a popularity measure of users. The same claim has been made on the basis that the number of followers is easily manipulated by link farming and spammer activities and the following may not directly reflect activities of the followers. Therefore, alternative popularity measures may be more useful.
In fact, previous studies proposed to rank Twitter users using, for example, the PageRank \cite{Cha2010ACWSM}, TwitterRank, i.e., a variant of the PageRank \cite{Weng2010ACM_ICWSDM}, TunkRank ({\it http://tunkrank.com/}), amount of activities received by the user including the number of retweets \cite{Leavitt2010WebEcology,Cha2010ACWSM,Kwak2010WWW} and mentions
\cite{Cha2010ACWSM}, and size of information cascades starting from a specified
user \cite{Bakshy2011ACM_ICWSDM}.

To suggest that the number of followers may not be an adequate measure for ranking users, we plot
the relationship between the number of followers and that of friends (i.e., those whom a user follows) in a scattergram
in Figure~\ref{fig:Japanese scattergram}(a). A point represents a
randomly sampled Japanese user that follows a specific Twitter user with $\approx 3.6\times 10^4$ followers.
The figure indicates that
some users possessing many followers have a small number of friends,
i.e., they follow a small number of other users. In contrast, other
users possessing similarly many followers follow many other users.
The number of the followers and that of friends are close (solid
diagonal line in Figure~\ref{fig:Japanese scattergram}(a)) for the
latter type of users.
The Twitter imposes that a user cannot have
much more friends than followers.
Then, it is not surprising that we do not find
users far off below the diagonal. However, we emphasize that many users are located near a vertical line corresponding to a small number of friends or near the diagonal.
Therefore, it may be beneficial to use the number of friends in addition to that of followers to assess the popularity of users. Furthermore, we
find few users with many followers and an intermediate number of friends.
We are interested in the generality and implications of this result.

In the present study, 
we sample local (i.e., egocentric) networks of Twitter users separately for some major countries.
We quantify differences between local followership networks around two types of users using five quantities and the PageRank. Based on the results we argue that, although the two types of users have similar numbers of followers, they are distinct in the number of friends in some countries. 
We propose that users with many followers are popular only when they follow a small number of other users.
Our preliminary results have been published in the form of a short conference proceeding \cite{SaitoMasuda2013ASONAM}.

\section*{Materials and methods}

\subsection*{Data sets}

Users of Twitter can read other users' tweets by registering their accounts, i.e., by following them. We refer to the directed network of
users in which a link is directed from the follower to the followee as the followership network.

We collected all the data between September 29, 2012 and January 11, 2013.
We used Twitter representational state transfer application programming interface (API) \cite{Russell2011book_mining} to collect data.
In particular, we acquired users' properties including the number of followers ({\bf followers\_count}), the number of friends ({\bf friends\_count}, i.e., 
the number of users that a user follows), and the language ({\bf lang}). The operating institution of Twitter allowed general users including us to collect the Twitter users' network at a limited speed.
We registered an application of Twitter as a developer
and authenticated the application by the OAuth 2.0 protocol
to use the so-called {\bf users/lookup}, {\bf followers/ids}, and {\bf friends/ids} resources.
The {\bf followers/ids} and {\bf friends/ids} resources return error when the targeted users protect their tweets
and are not followed by our test account. 
To acquire IDs of friends and followers of such protected users, we would have to beg them to accept our following.
Therefore, we excluded the protected users, which accounted for 1--10\% of the entire users, from the following analysis.

The correlation between the number of followers and that of
  friends is shown in a previous study \cite{Java2007WebKDD}, but not
  as strong as that implied in Figure~\ref{fig:Japanese
    scattergram}(a). In 2007, Twitter was much less known than it is
  now. Therefore, their data and contemporary data including ours can
  be different in demography. In particular, Twitter is now used in
  various countries, and its usage may depend on countries. Therefore,
  we decided to sample local networks centered around
  users with many followers 
separately for some major countries, where the classification
  is based on the language and location of the users. We
focused on users registering either of the seven
languages, i.e., English, Spanish, Japanese, Portuguese, Russian,
Korean, and French. These seven languages are used by many users such
that they are amenable to language-wise analysis.  The local network
of a user would be also homogeneous in terms of the language because
users tend to be connected with other users registering the same
language \cite{Takhteyev2012SocNet}.

\subsection*{Neighbor sampling}

We are concerned with local networks of users with relatively many followers. 
We sampled such users by the two methods called the neighbor sampling and random sampling defined as follows. 

In the neighbor sampling, we first selected seed users and then sampled followers of the seed users. 
It should be noted that we are not interested in the seed users. 
We defined users with many followers, as identified by the ``twitaholic'' website ({\it http://twitaholic.com/}), as seed users,
to realize a large sample size.
To this end, for seven countries where the corresponding languages were spoken as the dominant official language
(i.e., US, Spain, Japan, Brazil, Russia, Korea, and France),
we identified users whose residence location property contained the name of the city with the largest population in the country.
Then, for each of the seven countries, we selected three users as seeds under the condition that they were not accounts created by an organization or company
and that the three users had the largest number of followers among those having less than $5\times10^5$ followers in each country.
We excluded users with more than $5\times10^5$ followers from the seeds. This is because we had to collect the IDs of all of their followers to implement the random sampling explained in the following, and the Twitter's API did not allow us to collect users' data at a sufficiently high speed.

After determining 21 seed users in total, we acquired the IDs of the seeds' all followers.
The speed restriction of the API made it difficult for us to collect the local networks of all the seeds' followers.
Therefore, for each seed user, we randomly selected $5\times10^4$ users out of all the followers. It should be noted that $1.5\times 10^5$ users were sampled for each of the seed user's language.
Finally, homophily with respect to the language implies that the seeds' followers tend to register the same language as that of the seed user \cite{Takhteyev2012SocNet}. Because we will separately analyze users for different language groups, for each seed user, we filtered the already selected $5\times 10^4$ followers such that we discarded those registering a different language from that used by the seed user.
When the following analysis required local networks of the selected users,
we also acquired the information about the local networks of these users.

\subsection*{Random sampling}

In the random sampling, we randomly created $1.5\times10^6$ IDs as uniformly and independently distributed integers between 12 (corresponding to the first user) and the maximum ID value
among those of the seeds' followers identified by the neighbor sampling.
Then, we sifted out the users registering either of the seven target languages.

We used the two sampling methods for the following reasons. 
First, with the neighbor sampling, a sampled user tended to have much more followers than with the random sampling.
Therefore, the neighbor sampling allows us to investigate the statistics of users having many followers as compared to the random sampling does.
It should be noted that this empirical fact is independent of the theoretical fact that
the users sampled under the condition that they follow somebody have a larger number of friends than with the random sampling on average in heterogeneous networks.
Second, with the neighbor sampling, properties of the sampled users may be correlated
because a large fraction of them follows the same seed user. 
The random sampling method does not suffer from such correlation. 
Third,
the users collected by the neighbor sampling may be biased in the sense that
seeds are often popular and followed by new users. In contrast,
the random sampling approximates the unbiased random sampling.

\subsection*{Sample sizes}

We did not filter users according to their activities except that the
IDs banned by Twitter or deleted by users were neglected.
Our samples may contain spammers. Nevertheless, at least the users collected by the neighbor sampling were mostly not spammers
because they followed a celebrity user by definition. Up to our manual inspections, most users collected by either sampling method were not spammers. 

The sample sizes for the different sampling methods and languages are summarized in Table~\ref{tab:number of users}. 
In the neighbor sampling, we sampled $1.5\times 10^5$ users for each language and discarded those using a language different from the seed user's one. 
Among the $1.5\times 10^5$ users that followed a seed user registering English, $78.9$\% of the users also registered English.
Table~\ref{tab:number of users} indicates that this fraction depends much on the language, with the largest and smallest values being $86.3$\% for Spanish and $32.2$\% for Korean, respectively. Such a language dependence exists probably 
because some languages including English and Spanish are spoken by many users and because some seeds are globally famous and other seeds are not.
In the random sampling, the number of the users summed over the seven languages is equal to 913,426. Therefore, $[1-913,426/(1.5\times 10^6)]\times 100=39.1$\% of the users were discarded because the language was not the same as the seed user's one or the sampled ID was invalid.

\section*{Results}

\subsection*{Overview of the Results}

The present section is organized as follows. In the first three subsections, we define two user types, referred to by type 1 and type 2.
In the subsequent subsections ``Local link reciprocity of type 1 and 2 users''
through ``Abundance of type 2-like users among followers'',
we compare type 1 and type 2 users by examining five quantities derived from their local networks. With the API, the acquisition
of the information about the local networks of users is costly in terms of time. Therefore, we decided to use ten users of each type and language for the analysis in these subsections. In the subsection
``PageRank of the two types of users'', we assess the PageRank of the two types of users.
In Table~\ref{tab:summary},
we summarize the results shown in
subsections ``Local link reciprocity of type 1 and 2 users''
through ``PageRank of the two types of users''.

\subsection*{Distribution of the number of followers and friends}

\begin{figure}[htbp]
\begin{center}
\includegraphics[width =8.0cm]{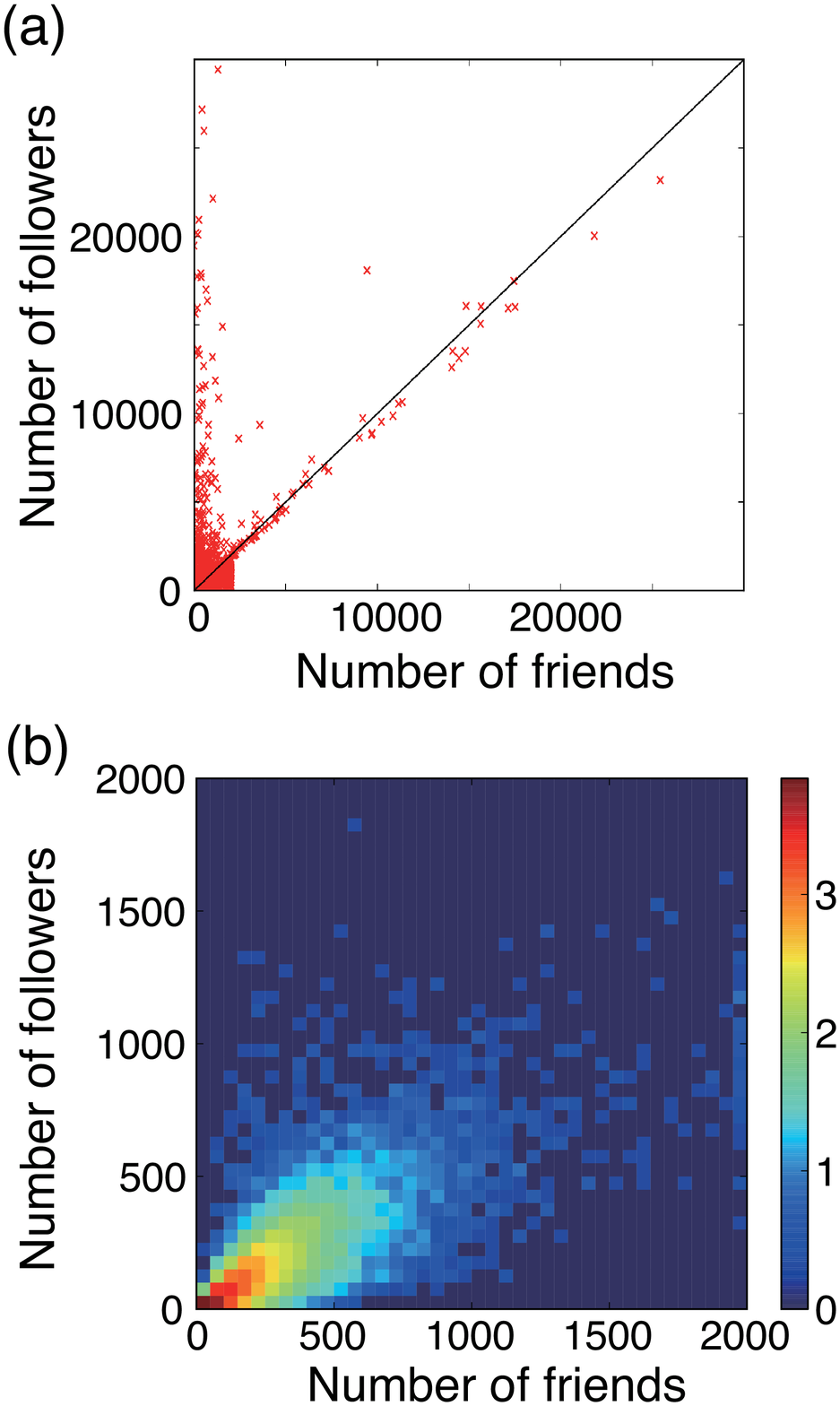}
\caption{(a) Relationship between the
number of friends and that of followers for the 34075 Japanese users following
a specific Japanese Twitter user.
(b) Density plot of the number of friends and that of followers for the users that are shown in (a) and have with less than $2\times 10^3$ friends and followers.}
\label{fig:Japanese scattergram}
\end{center}
\end{figure}

First of all, Figure~\ref{fig:Japanese scattergram}(a)
indicates that a small fraction of users has a large number of followers or friends. In quantitative terms, 
the distribution of the number of followers and that of friends obey
long-tailed distributions. This is the case in networks in various domains including Twitter's social networks \cite{Ghosh2012WWW,Weng2010ACM_ICWSDM,Kwak2010WWW}. 

In the present paper, we focus on a different property evident in
Figure~\ref{fig:Japanese scattergram}(a), which is the joint distribution of the number of followers and friends. As briefly explained in Introduction, the users possessing many followers seem to be classified into two types according to the number of friends. In particular, some users have equally large numbers of followers and friends. To assess the generality of this observation, we show in
Figure~\ref{fig:Japanese scattergram}(b) the density plot that magnifies Figure~\ref{fig:Japanese scattergram}(a). We use the density plot because there are many users with small numbers of followers and friends. In this region, there is no system restriction on the number of followers and that of friends; any user is allowed to possess up to $2\times 10^3$ followers and friends. Figure~\ref{fig:Japanese scattergram}(b) indicates that many users are concentrated on the diagonal. This result is consistent with that for large numbers of followers and friends shown in Figure~\ref{fig:Japanese scattergram}(a).

\subsection*{Identification of users having approximately many followers and friends}

In Figure~\ref{fig:Japanese scattergram}, we showed that some users following a Japanese popular user have similar $k^{\rm in}$ (i.e., number of followers) and $k^{\rm out}$ (i.e., number of friends) values.
To generalize and scrutinize this observation, we measure two quantities for each language group.
First, we define the degree ratio by
\begin{equation}
r = \left\langle\frac{\min(k^{\rm in}, k^{\rm out})}{\max(k^{\rm in}, k^{\rm out})}\right\rangle,
\end{equation}
where $\left\langle \cdot \right\rangle$ represents the average over the users in a language group.
If $k^{\rm in}$ and $k^{\rm out}$ are close for many users, $r$ is large.
Second, we define the diagonal fraction, denoted by $d$, as the fraction of users that satisfy
\begin{equation}
k^{\rm out}/1.1 \leq k^{\rm in} \leq 1.1\times k^{\rm out}.
\end{equation}
The factor $1.1$ originates from the fact that the operating institution of Twitter does not seem to allow users with $k^{\rm out} \ge 2\times 10^3$ to have $k^{\rm out} \ge 1.1\times k^{\rm in}$ friends.

Both $r$ and $d$ range between 0 and 1.
The $r$ and $d$ values may be strongly affected by users having small $k^{\rm in}$ and $k^{\rm out}$ values, 
which occupy the majority owing to the long-tailed distributions of $k^{\rm in}$ and $k^{\rm out}$ \cite{Ghosh2012WWW,Weng2010ACM_ICWSDM,Kwak2010WWW,Welch2011WSDM}. 
Because in this study we focus on properties of users having relatively many friends and followers,
we restrict ourselves to the users satisfying $k^{\rm in}, k^{\rm out} >100$ or $k^{\rm in}, k^{\rm out} > 2000$.
	
The $r$ and $d$ values for the different sampling methods, language groups, and threshold degrees (i.e., 100 or 2000) are shown in Table~\ref{tab:r and d}.
Regardless of the sampling method and threshold degree, $r$ and $d$ are large for the Japanese, Russian, and Korean groups,
intermediate for the English group, and small for the Spanish, Portuguese, and French groups. 
Therefore, the observation that many users have similar in-degree and out-degree, as shown in Figure~\ref{fig:Japanese scattergram} for Japanese users, 
is eminent for Japanese, Russian, and Korean among the seven languages.

\subsection*{Definition of type 1 and 2 user}

\begin{figure}[htbp]
\begin{center}
\includegraphics[width = 8.5cm, bb=0 0 800 400]{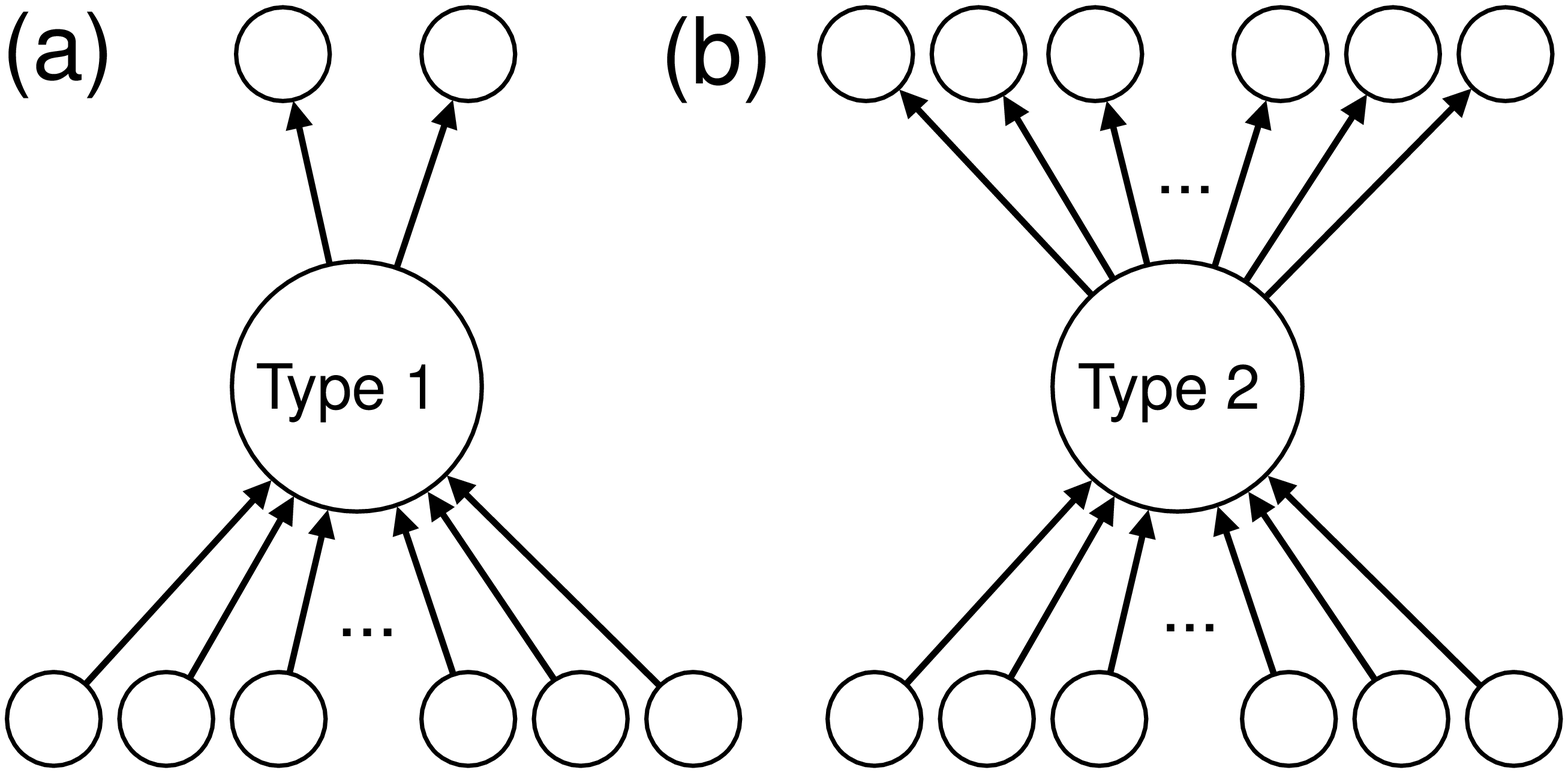}
\caption{Schematic of the two types of users with many followers. (a) Type 1 user. (b) Type 2 user.}
\label{fig:2 types}
\end{center}
\end{figure}

Our main hypothesis is that the quality of the follow may be different between users with large $k^{\rm out}$ and those with small $k^{\rm out}$
even if the users enjoy equally many followers (i.e., large $k^{\rm in}$). 
To investigate this issue on the basis of the followership network, we classify users with many followers into two types as follows (Figure~\ref{fig:2 types}).
We define type 1 users as those satisfying 2500 $\le k^{\rm in} \le$ 7500 and $k^{\rm out} \le$ 500.
Type 1 users are followed by many users and do not follow many others. We define type 2 users as those
 satisfying $k^{\rm out}/1.1 \leq k^{\rm in} \leq 1.1\times k^{\rm out}$ and
$5000 \leq k^{\rm in} + k^{\rm out}\leq 15000$.
Type 2 users are followed by many users and follow many others.
Many users are located near the diagonal in Figure~\ref{fig:Japanese scattergram} partly because a user with $k^{\rm out} \ge 2\times 10^3$ cannot own $k^{\rm out} \ge 1.1\times k^{\rm in}$ friends, as mentioned before.
Nevertheless, we are interested in the behavior of type 2 users.

The in-degree $k^{\rm in}$ of type 2 users is distributed on roughly the same range as $k^{\rm in}$ of type 1 users (i.e., 2500 $\le k^{\rm in}\le 7500$). Therefore, type 1 and 2 users are indifferent in terms of $k^{\rm in}$.
We may be able to reveal the difference between the two types of users by inspecting contents of the tweets and other
activities of these users (e.g., tweet and retweet rates). In the following, we take a complementary, purely network-based approach.

\subsection*{Local link reciprocity of type 1 and 2 users}\label{sub:local link reciprocity}

First, we examine the so-called
local link reciprocity (reciprocity for short) of a user
defined as the number of the focal user's friends
that follow back the focal user, divided by $k^{\rm out}$ of the focal user.
The local link reciprocity takes a value between 0 and 1.
We hypothesize that type 2 users, not type 1 users,
have much larger reciprocity values
 because type 2 users would follow back their followers to maintain reciprocal links.
By definition, $k^{\rm out}$ values 
for type 1 and 2 users are very different. Therefore,
the reverse definition of the reciprocity,
i.e., the number of the focal user's followers that a focal type 1 or 2
user follows back, divided by $k^{\rm in}$ of the focal user, does not serve to 
examine the difference between type 1 and 2 users.
This is because the upper bound of the reversed reciprocity
is much smaller for type 1 users than type 2 users.

For each language group, the mean and standard deviation of the reciprocity of the ten randomly selected users of type 1 or 2 are shown in Table~\ref{tab:local link reciprocity}.
The table indicates that type 2 users have significantly larger reciprocity than type 1 users,
at least for the Japanese, Russian, and Korean groups, for which the distinction between the type 1 and 2 users is clear (Table~\ref{tab:r and d}).
It should be noted that approximately 80~\% of links in Twitter are reciprocal \cite{Weng2010ACM_ICWSDM} (also see \cite{Hopcroft2011CIKM} for link reciprocity in the Twitter social network; but also see \cite{Cha2010ACWSM}).
This is consistent with the results shown in Table~\ref{tab:local link reciprocity}, in which the reciprocity values are generally large.

\subsection*{Out-degree of those following a type 1 or 2 user}

\begin{figure}[htbp]
\begin{center}
\includegraphics[width = 8.0cm, bb=0 0 800 400]{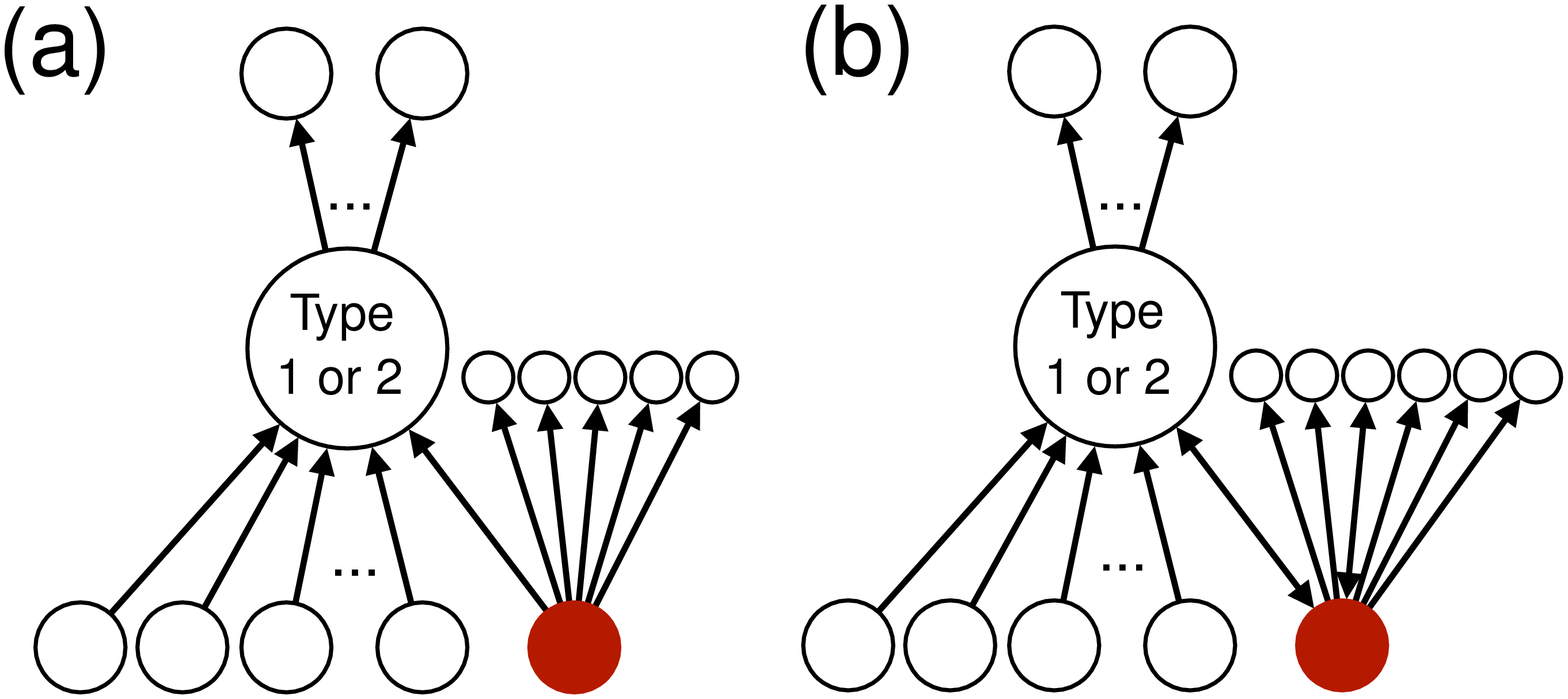}
\caption{(a) Out-degree of those following a type 1 or 2 user. It is equal to 6 for the user shown by the filled circle. (b) Follower's reciprocity. It is equal to $2/7$ for the user shown by the filled circle.}
\label{fig:out-degree and reciprocity}
\end{center}
\end{figure}

Second, we examine $k^{\rm out}$ (i.e., number of friends) for those following a type 1 or 2 user (Figure~\ref{fig:out-degree and reciprocity}(a)). If $k^{\rm out}$ is large, 
the follow that a type 1 or 2 user receives may not be valuable because the amount of time that a follower spends on looking at others' tweets would be inversely proportional 
to $k^{\rm out}$ to the first-order approximation.

\begin{figure}[htbp]
\begin{center}
\includegraphics[width = 9.0cm]{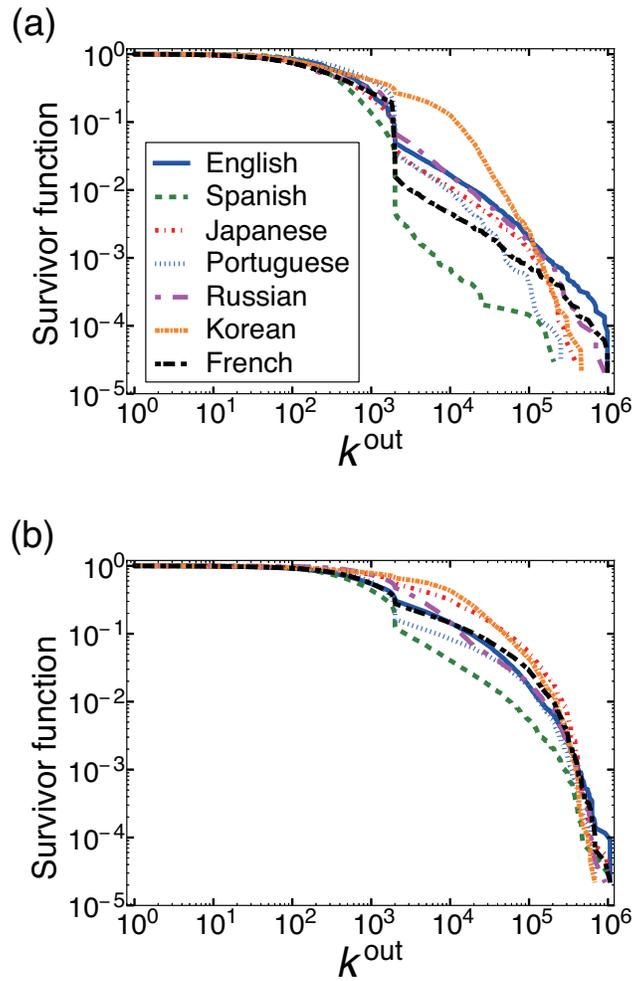}
\caption{Survivor function of the number of friends (i.e., $k^{\rm out}$)
for the followers of a (a) type 1 user and (b) type 2 user.
The sudden drop at $k^{\rm out}=2000$ is caused by the system restriction
that users having 
more than 2000 friends are disallowed to possess $k^{\rm out} \ge 1.1\times k^{\rm in}$ friends.}
\label{fig:kout followers}
\end{center}
\end{figure}

For those that follow any of the ten selected type 1 or 2 users of each language, the survivor functions of $k^{\rm out}$ (i.e., fraction of users whose
$k^{\rm out}$ is larger than a specified value) are shown in Figure~\ref{fig:kout followers}(a) and \ref{fig:kout followers}(b) for the type 1 and 2 user, respectively.
Figure~\ref{fig:kout followers} indicates that a follower of a type 2 user tends to have larger $k^{\rm out}$ than a follower of a type 1 user on average. For the Japanese, Russian, and Korean groups, the mean $\pm$ standard deviation, rounded to integer values, is equal to
1,125 $\pm$ 7,193 for type 1 and 20,070 $\pm$ 48,849 for type 2, 1,526 $\pm$ 11,435 for type 1 and 9,068 $\pm$ 31,711 for type 2, 4,119 $\pm$ 11,316 for type 1 and 20,114 $\pm$ 40,424 for type 2, respectively.

Because $k^{\rm out}$ obeys relatively long-tailed distributions (Figure~\ref{fig:kout followers}), the comparison of the mean values is insufficient. Therefore, we quantify the classification performance of the follower's $k^{\rm out}$ by using the receiver operating characteristic curve (ROC)
based on the two distributions of $k^{\rm out}$ for each language \cite{Tuffery2011book}. 
The ROC is the trajectory of the false positive (i.e., fraction of type 2 users that are mistakenly judged as type 1 
on the basis of $k^{\rm out}$) and the true positive (i.e., fraction of type 1 users correctly judged as type 1 with the same threshold),
when the threshold for classification is varied. 
The area under the curve (AUC) of the ROC falls between 0.5 and 1. When AUC is large, the two distributions are well separated such that
users are accurately judged as type 1 or 2.
The values of AUC for different language groups are shown in Table~\ref{tab:AUC}. 
The AUC is larger for the Japanese, Russian, and Korean groups than for the other four groups.
It should be noted that for the Japanese, Russian, and Korean groups, the type 1 and type 2 users are more clearly distinguished than for the other groups (Table~\ref{tab:r and d}).

\subsection*{Follower's reciprocity}

Third, we measure the number of reciprocal links owned by a follower of a type 1 or 2 user, divided by $k^{\rm out}$ for this follower (Figure~\ref{fig:out-degree and reciprocity}(b)).
We call the ratio the follower's reciprocity,
which ranges between 0 and 1. If the follower's reciprocity is large, the follow that a type 1 or 2 user receives may not be valuable in the sense that the follower
easily establishes reciprocal links with others, perhaps to advertise themselves  \cite{Ghosh2012WWW} or mutually connect with close friends.

To calculate the follower's reciprocity and also the fourth quantity $C_{i}$ described below,
we have to acquire IDs of the followers and friends for each user following a type 1 or 2 user.
This operation requires much time because we can call API resources a limited number of times per hour.
Therefore, we calculate the quantity of interest (follower's reciprocity or $C_{i}$) for randomly selected 100 users following each type 1 or 2 user.

We found that followers of type 2 users have larger follower's reciprocity values than followers of type 1 users on average. This holds true in particular for the Japanese
(0.434 $\pm$ 0.250 for type 1 versus 0.762 $\pm$ 0.224 for type 2, where the mean and standard deviation are calculated on the basis of all the users that follow any of the ten randomly selected type 1 or 2 users),
Russian (0.231 $\pm$ 0.287 for type 1 versus 0.703 $\pm$ 0.266 for type 2), and Korean (0.491 $\pm$ 0.352 for type 1 versus 0.846 $\pm$ 0.206 for type 2) groups.
Because the follower's reciprocity in fact obeys a rather long tailed distribution, we calculate the AUC for the follower's reciprocity.
The AUC values for the seven language groups are shown in Table~\ref{tab:AUC}. The AUC is relatively large such that the follower's reciprocity is effective at distinguishing between type 1 and 2 users.

\subsection*{Local clustering coefficient}

Fourth, we examine the local clustering coefficient \cite{Newman2003SiamRev,Newman2010book}, denoted by $C_i$ for type 1 or 2 user labeled $i$, which is the density of triangles including user $i$.
For a type 1 or 2 user $i$ having in-degree $k_i^{\rm in}$, there can be at most $k_i^{\rm in}(k_i^{\rm in} - 1)/2$ triangles that include user $i$,
whereby we impose that two followers of $i$ are connected by reciprocal links to be qualified as a triangle including $i$.
We define
\begin{equation}
C_i = \frac{\mbox{Number of triangles containing user } i}
{k_i^{\rm in}(k_i^{\rm in} - 1)/2}.
\end{equation}
By definition, $C_i$ ranges between 0 and 1.
Because the Twitter followership network has a large global clustering coefficient \cite{RomeroKleinberg2010conf},
a considerable portion of users would have large $C_i$.
If $C_i$ is large, the follow that a type 1 or 2 user $i$ receives may be not as valuable as otherwise
because the user is likely to be followed by many similar users, where the similarity is implicit in reciprocal links between the followers.

As shown in Table~\ref{tab:local clustering coefficient}, $C_{i}$ is significantly larger for type 2 users than type 1 users except for the Portuguese group.
It should be noted that the difference is prominent for the Japanese, Russian, and Korean groups, for which the distinction between the type 1 and type 2 users are clear.

\subsection*{Abundance of type 2-like users among followers}\label{sub:type 2prime}

Fifth, we define the fraction of type 2-like users among the followers. It should be noted that $k^{\rm out}$ of the followers (second quantity that we have investigated) and the follower's reciprocity (third quantity) also capture the tendency that users following a type 1 or 2 user resemble type 2 users to some extent. Here we define a more direct measure called the fraction of type $2^{\prime}$ users
as the fraction of followers of a type 1 or 2 user satisfying $k^{\rm out}/1.1 \leq k^{\rm in} \leq 1.1\times k^{\rm out}$.
Similar to the definition of $d$, we exclude the followers with $k^{\rm in}$ and $k^{\rm out}$ values smaller than a prescribed threshold
from the calculation of the fraction of type $2^{\prime}$ users. 
The analysis of the four quantities carried out above suggests that the follow that a type 2 user receives is probably less valuable than that a type 1 user receives.
If we accept this assumption, a large fraction of type $2^{\prime}$ users among the followers of type 2 users as compared to among the followers of type 1 users would lend
another support to our claim that the follow that a type 2 user receives is not as valuable as that a type 1 user receives.
For each user type and language, we calculate the mean and standard deviation of the fraction of type $2^{\prime}$ users on the basis of the ten randomly selected users.

The results with the threshold equal to 100 (i.e., followers having $k^{\rm in}$, $k^{\rm out}$ $\le $ 100 are excluded from the calculation of the fraction of type $2^{\prime}$ users) and 2000 are shown in Table~\ref{tab:type 2prime}. The table indicates that type 2 users are significantly more likely to be followed by type $2^{\prime}$ users than type 1 users are. 
This tendency is stronger for the Japanese, Russian, and Korean groups than the other four language groups.

\subsection*{PageRank}\label{sub:PageRank}

In this subsection, we estimate the PageRank of type 1 and type 2 users. It should be noted that all the quantities measured in the previous sections are local ones, whereas the PageRank quantifies global importance of a node in
directed networks \cite{Brin1998conf,Langville2006book}.
In fact, the PageRank and its variants have been used for ranking users in Twitter social networks \cite{Weng2010ACM_ICWSDM,Hopcroft2011CIKM,Welch2011WSDM,Chang2013WSDM}. 
By definition, the PageRank of a user would be small if the user's follower has a large $k^{\rm out}$(i.e., number of friends). 
Therefore, we expect that a type 1 user in general has a larger PageRank value than a type 2 user with the same number of followers.
The PageRank of a node is proportional to the frequency with which a random walker visits the node. The walker is defined to move to one of downstream neighbors with the equal probability $(1-q)/k^{\rm out}$ such that the total probability of such an ordinary random walk is equal to $1-q$. With the remaining probability $q$, the walker jumps to an arbitrary node with the equal probability, which is the so-called teleportation. Although the PageRank is often strongly correlated with $k^{\rm in}$ \cite{Fortunato2008LNCS,Ghoshal2011NatComm},
it is not always the case \cite{Donato2004EPJB,MasudaOhtsuki2009NewJPhys}. For Twitter networks, it was reported that $k^{\rm in}$ (i.e., number of followers) and the PageRank were strongly correlated \cite{Kwak2010WWW}.

Because the exact calculation of the PageRank requires the full information about the connectivity of the network,
we approximate the PageRank by emulating the random walk.
We first select a user with the equal probability from the set of users. The random walk starts from the selected user.
We selected the initial position of the random walk from the set of Japanese users collected by the random sampling.
We confined ourselves to Japanese users because the distinction between type 1 and 2 users is clear for them. 
Second, we move to a friend of the selected user with the equal probability $1/k^{\rm out}$.
Third, we repeat the same random hopping ten times.
If the walker hits a user without any follower before hopping ten times, we terminate the random walk.
Finally, we redraw a starting user without replacement and carry out the ten-step random walk for 1500 randomly selected initial nodes. Stopping the random walk after ten steps corresponds to the teleportation 
with probability $q = 1/11$. This value is comparable with the conventional teleportation probability $q = 0.15$ \cite{Brin1998conf,Langville2006book}.
The probability that the walker hits a given type 1 or 2 user is very small.
To enhance the probability that the walker hits any of type 1 or 2 users, we increased the number of type 1 users and that of type 2 users as follows.
First, we focused on type 1 and 2 Japanese users identified by the neighbor sampling because it is much rarer to find a type 1 or 2 user with the random sampling. 
Second, we added two Japanese seed users. We scanned all followers of the two seed users to find new type 1 and 2 users employed as additional targets of the random walk.

Because the PageRank is usually correlated with $k^{\rm in}$, we counted the number of visits to type 1 or 2 users for each of the four groups defined by different $k^{\rm in}$ ranges (Table~\ref{tab:PageRank}).
For each degree group, the walker visits type 1 users more frequently than type 2 users. Therefore, we conclude that
type 1 users are more important than type 2 users in terms of the PageRank.

\section*{Discussion and Conclusions}

By measuring several network-based quantities, we showed that type 1 and 2 users had different network properties although they had comparably many followers. On average, type 1 users, defined by a small number of friends, are characterized by less reciprocal links, possession of followers with less reciprocal links and less friends, and larger PageRank values, than type 2 users.
We also found that the difference between the type 1 and 2 users is more clear cut for Japanese, Russian, and Korean users than for English, Spanish, Portuguese, and French users. On the basis of these results, we propose that a follow that a type 1 user receives is more valuable than one that a type 2 user receives.
Announcing that a given user is type 2 user may serve to maintain social etiquette in the Twitter blogosphere.

A fraction of the sampled type 1 and 2 users was spammers, organizational accounts, and bots. However, we manually inspected the sampled users to
find that few of them were spammer-like accounts. This was in particular the case for the Japanese and Spanish users. Therefore, the effects of the spammer-type accounts on the present results are considered to be small.

User IDs suspected of organized link farming activities may follow other users and anticipate that they are followed back.
Such users may be the so-called social capitalists, 
who aim to promote their legitimate contents to be broadcast to wide audience \cite{Ghosh2012WWW}. They tend to exchange reciprocal links with others
and are densely connected with each other. 
Similar to social capitalists, spam followers also tend to have high reciprocity. These behavioral properties of social capitalists are 
consistent with the high reciprocity and homophily of type 2 users found in the present study.
However, analysis of the intention and behavior of the type 2 users is beyond the scope of the present study; we analyzed the followership networks but not the contents or propagation of tweets.
It should be also noted that, unlike Ghosh et al. \cite{Ghosh2012WWW}, we did not look at connectivity of users to spams.
Type 2 users may exchange links as a part of link farming activities, spam activities, or just to assure mutual friendship.

Ghosh et al. cite celebrities and popular bloggers as examples of social capitalists \cite{Ghosh2012WWW}. However, our manual inspection of  the users' profiles suggests that more celebrities and popular bloggers are found among type 1 rather than type 2 users. They also conclude that social capitalists and spammers are influencers \cite{Ghosh2012WWW}.
In contrast, our type 2 users would have much smaller influences in terms of the PageRank than type 1 users. Although the reason for this discrepancy is unclear, our main claim is that we can classify seemingly influential (i.e., having large number of followers) users into rather distinct two types. Social capitalists identified by Ghosh et al. \cite{Ghosh2012WWW} may be a mixture of type 1 and 2 users.
To subcategorize the social capitalists into type 1 and type 2 -like classes by incorporating the information about tweets and connectivity to spams is warranted for future work.

The number of followers and that of friends were very close for most users in a previous report \cite{Weng2010ACM_ICWSDM}. The results are inconsistent with ours; we found that the proximity depends on users (Figure~\ref{fig:Japanese scattergram}) and the language (Table~\ref{tab:r and d}). Although unclear, type 1 users were not found in the previous study  \cite{Weng2010ACM_ICWSDM} perhaps because they mainly investigated English speaking users.

Weng et al. proposed the TwitterRank to rank users \cite{Weng2010ACM_ICWSDM}. The TwitterRank is different from the PageRank because in the former the walker tends to 
transit to a friend that is similar to the user and tweets many times on each topic. The TunkRank is another variant of the PageRank in which the retweet probability is taken into account in determining the transition probability ({\it http://tunkrank.com/}).
In the present work, we used the original PageRank without taking these non-network features into account. Our aim was to extract the information about the value of users only on the basis of the network structure. Better characterizing different types of users by combining the present method with users' activities is an obvious future question. Use of networks other than the followership network induced by Twitter data, such as the networks defined by retweets \cite{Hopcroft2011CIKM,Welch2011WSDM,Chang2013WSDM}, may be promising to this end.

Web Ecology project measures the influence of the user on the basis of the activities received by the user, which include the number of retweets divided by that of tweets \cite{Leavitt2010WebEcology}.
Our results are in line with this definition because a network equivalent of their measure is given by $k^{\rm in}$/$k^{\rm out}$, which is much larger than unity for type 1 users and approximately equal to unity for type 2 users.

\section*{Acknowledgments}
We thank Katsumi Sakata for discussing ideas leading to this project, and Mitsuhiro Nakamura, Ryosuke Nishi, and Taro Takaguchi for critical reading of the manuscript.


\newpage
\clearpage

\clearpage

\clearpage

\clearpage

\clearpage

\begin{table}[htbp]
  	\begin{center}
	\caption{Number of users sampled by the neighbor and random sampling methods.}
\label{tab:number of users}
  	\begin{tabular}{|c|c|c|}\hline
    	Language & Neighbor sampling & Random sampling\\ \hline \hline
    	English & 118,316 & 638,122\\
    	Spanish & 129,415 & 126,350\\
   	Japanese & 113,140 & 44,204\\
    	Portuguese & 95,211 & 43,353\\
    	Russian & 70,354 & 24,940\\
    	Korean & 48,367 & 13,636\\
    	French & 51,571 & 22,821\\ \hline
    \end{tabular}
  \end{center}
\end{table}

\newpage

\begin{table}[htbp]
  \begin{center} 
          \caption{Summary of the results.} 
\label{tab:summary}
          \begin{tabular}{|c|c|c|}
            \hline
            Property&Type 1&Type 2\\ \hline \hline
            Local link reciprocity & small & large\\
            Follower's $k^{\rm out}$ & small & large\\
            Follower's reciprocity & small & large\\
            Local clustering coefficient & small & large\\
            Fraction of type $2^{\prime}$ users & small & large\\
            PageRank & large & small\\\hline
            \hline
          \end{tabular}
  \end{center}
\end{table}

\newpage

\begin{table}[htbp]
  \begin{center}
         \caption{Degree ratio ($r$) and the diagonal fraction ($d$) for the users satisfying $k^{\rm in}, k^{\rm out} > 100$ (values left to the slash) and 2000 (values right to the slash).}
\label{tab:r and d}
          \begin{tabular}{|c|c|c|c|c|}
            \hline
            Language & $r$(neighbor) & $r$(random) & $d$(neighbor) & $d$(random) \\ \hline \hline
            English & 0.299/0.532 &0.429/0.415 &0.031/0.209 &0.080/0.180\\
            Spanish & 0.360/0.257&0.395/0.399&0.031/0.050&0.059/0.179\\
            Japanese & 0.585/0.635&0.695/0.722&0.115/0.333&0.250/0.473\\
            Portuguese & 0.232/0.315 &0.386/0.342&0.013/0.049&0.051/0.090\\
            Russian & 0.408/0.759&0.409/0.627&0.091/0.517&0.074/0.500\\
            Korean & 0.439/0.752&0.598/0.824&0.072/0.548&0.218/0.685\\
            French & 0.313/0.464&0.379/0.238&0.028/0.169&0.048/0.036\\ \hline
            \hline
    \end{tabular}
  \end{center}
\end{table}

\newpage

\begin{table}[htbp]
  \begin{center} 
          \caption{Local link reciprocity for different language groups.} 
\label{tab:local link reciprocity}
          \begin{tabular}{|c|c|c|}
            \hline
            Language&Type 1&Type 2\\ \hline \hline
            English&0.364$\pm$0.240&0.656$\pm$0.230\\
            Spanish&0.478$\pm$0.181&0.669$\pm$0.192\\
            Japanese&0.600$\pm$0.206&0.872$\pm$0.102\\
            Portuguese&0.280$\pm$0.234&0.420$\pm$0.233\\
            Russian&0.452$\pm$0.185&0.861$\pm$0.232\\
            Korean&0.648$\pm$0.214&0.884$\pm$0.069\\
            French&0.557$\pm$0.235&0.851$\pm$0.196\\\hline
            \hline
          \end{tabular}
  \end{center}
\end{table}

\newpage

\begin{table}[htb]
 \begin{center}
  \caption{AUC values for the follower's $k^{\rm out}$ and the follower' reciprocity.}
\label{tab:AUC}
  \begin{tabular}{|c|c|c|}\hline
    Language &Follower's& Follower's\\
    &$k^{\rm out}$&reciprocity\\ \hline \hline
    English & 0.680 & 0.815\\
    Spanish & 0.704 & 0.740\\
    Japanese & 0.831 & 0.838\\
    Portuguese & 0.628 & 0.681\\
    Russian & 0.819 & 0.874\\
    Korean & 0.748 & 0.796\\
    French & 0.721 & 0.883\\ \hline 
  \end{tabular}
 \end{center}
\end{table}

\newpage

\begin{table}[htbp]
  \begin{center} 
          \caption{Local clustering coefficient. The mean and standard deviation are calculated on the basis of ten randomly selected users of each type and language.} 
\label{tab:local clustering coefficient}
          \begin{tabular}{|c|c|c|}
            \hline
            Language&Type 1&Type 2\\ \hline \hline
            English&0.0036$\pm$0.0087&0.0293$\pm$0.0275\\
            Spanish&0.0017$\pm$0.0016&0.0098$\pm$0.0077\\
            Japanese&0.0039$\pm$0.0039&0.1334$\pm$0.0875\\
            Portuguese&0.0025$\pm$0.0034&0.0214$\pm$0.0417\\
            Russian&0.0086$\pm$0.0110&0.0919$\pm$0.0359\\
            Korean&0.0988$\pm$0.1505&0.3648$\pm$0.2197\\
            French&0.0021$\pm$0.0027&0.0419$\pm$0.0341\\\hline
            \hline
          \end{tabular} 
  \end{center}
\end{table}

\newpage

\begin{table}[h]
  \begin{center} 
          \caption{Fraction of type $2^{\prime}$ users for different user types and languages.}
\label{tab:type 2prime}
          \begin{tabular}{|c|c|c|c|c|}
            \hline
            Language&Type 1&Type 2&Type 1&Type 2\\
            &(threshold$=$100)&(threshold$=$100)&(threshold$=$2000)&(threshold$=$2000)\\ \hline \hline
            English&0.055$\pm$0.046&0.244$\pm$0.112&0.212$\pm$0.115&0.402$\pm$0.099\\
            Spanish&0.022$\pm$0.008&0.123$\pm$0.045&0.057$\pm$0.059&0.357$\pm$0.078\\
            Japanese&0.122$\pm$0.049&0.486$\pm$0.141&0.326$\pm$0.121&0.674$\pm$0.065\\
            Portuguese&0.022$\pm$0.012&0.108$\pm$0.137&0.071$\pm$0.043&0.213$\pm$0.149\\
            Russian&0.091$\pm$0.089&0.397$\pm$0.055&0.279$\pm$0.174&0.603$\pm$0.045\\
            Korean&0.313$\pm$0.353&0.758$\pm$0.192&0.506$\pm$0.313&0.912$\pm$0.047\\
            French&0.034$\pm$0.025&0.248$\pm$0.109&0.132$\pm$0.065&0.449$\pm$0.136\\\hline
            \hline
          \end{tabular}
  \end{center}
\end{table}

\newpage

\begin{table}[h]
 \begin{center}
 \caption{Frequency that the random walker visits type 1 or 2 users. For each degree group defined by a distinct range of $k^{\rm in}$, we found less type 1 users than type 2 users by the neighbor sampling. 
 Therefore, we randomly sampled users from the set of type 2 users such that the number of type 2 users is equal to that of type 1 users (e.g., 941).}
\label{tab:PageRank}
  \begin{tabular}{|c|c|c|c|}\hline
    $k^{\rm in}$ & Number of users & Type 1 & Type 2\\ \hline \hline
    2500--7500 & 941 & 43 & 12\\
    7500--12500 & 224 & 16 & 4\\
    12500--17500 & 93 & 10 & 4\\ 
    17500--22500 & 62 & 10 & 3\\ \hline
  \end{tabular}
 \end{center}
\end{table}

\end{document}